\documentclass[%
 reprint,
 amsmath,amssymb,
 aps,
prd,
nofootinbib]{revtex4-2}

\usepackage{graphicx}
\usepackage{dcolumn}
\usepackage{bm}
\usepackage{xcolor}        
  
\usepackage[dvipsnames]{xcolor}
\usepackage[unicode]{hyperref}
\hypersetup{colorlinks=true, citecolor=MidnightBlue,
            linkcolor=MidnightBlue, urlcolor=MidnightBlue, linktocpage=true}
\usepackage{multirow}
\usepackage{array}

\newcommand{\lads}{L_{AdS_2}}

\begin{document}

\title{EFT Corrections to Majumdar-Papapetrou Black Holes}

\author{Soham Acharya}
\email{acharyasoham@iitgn.ac.in}
\author{Shuvayu Roy}%
\email{shuvayu.roy@iitgn.ac.in}
\author{Sudipta Sarkar}
\email{sudiptas@iitgn.ac.in}
\affiliation{Indian Institute of Technology, Gandhinagar, Gujarat 382055, India}%


\date{March 25, 2025}

\begin{abstract}
    Recent studies on extremal black holes within effective field theories (EFT) of gravity have revealed an intriguing phenomenon: tidal forces near the horizon experience significant enhancement due to EFT corrections, potentially leading to a breakdown of the EFT framework. In this work, we investigate this effect in a two-black-hole Majumdar-Papapetrou spacetime modified by four-derivative EFT correction terms. Our analysis shows that while the scaling exponents, which measure the strength of the tidal forces near the horizon, remain unchanged under EFT corrections in $D=4$, they decrease for higher dimensions, enhancing the near-horizon tidal forces. However, even in the presence of EFT corrections, these scaling exponents stay well-behaved and no such cases arise where the two-derivative contributions vanish and only the higher-derivative terms contribute. We find an expression for the EFT corrections to the scaling exponents till $D \le 10$ and demonstrate that the metric corrections can be structured such that only the near-horizon $AdS_2$ throat undergoes angle-dependent modifications, while the transverse $S^{D-2}$ sector remains unaffected. 
\end{abstract}

\maketitle


\section{Introduction}\label{sec:intro}

Einstein's theory of general relativity has been the reigning theory of gravitation for the last century since its inception. Besides leading to numerous theoretical predictions, it has also garnered massive observational support owing to the recent observations of gravitational waves \cite{LIGOScientific:2016aoc,KAGRA:2013rdx} and black holes \cite{EventHorizonTelescope:2019dse,EventHorizonTelescope:2024hpu,EventHorizonTelescope:2024rju}. It has also been a well-known result that the two-derivative theory of gravity, based on the Einstein-Hilbert action, would not suffice as the theory of gravitation across all energy scales, and any model of UV complete gravity would appear in the low-energy regimes as Einstein gravity with higher-derivative correction terms  \cite{Burgess:2003jk,Donoghue:2012zc}. Assuming that the high-energy degrees of freedom can be integrated out and that their effect manifests in the theory via these perturbative corrections only, one can treat these higher-derivative terms in an effective field theory (EFT) sense.

Black holes are a rich and diverse family of solutions to Einstein's equations with interesting thermodynamic properties. Such solutions come in a variety of forms, ranging from spherically symmetric vacuum and charged black holes to rotating solutions and their extensions in different dimensions. Just like their counterparts in two-derivative theories, one can have black hole solutions in theories with higher-derivative corrections \cite{Boulware:1985wk,Myers:2009ij,Svarc:2018coe}. In such solutions with higher-derivative corrections, one would expect that the higher-derivative terms would contribute as perturbative corrections to the leading-order effects in the spirit of the EFT approach. However, as recently observed in \cite{Horowitz:2022mly} and subsequent works, this is not always the case, and hence, the subtleties of these EFT-corrected black hole solutions demand further scrutiny.

While extremal black holes remain a very interesting class of solutions in their own right, recent explorations on extremal Kerr, Reissner-Nordstr\"om, and Kerr-Newman black holes \cite{Horowitz:2023xyl,Horowitz:2024dch} have uncovered the sensitivity of their near-horizon physics to EFT corrections. It has been shown that perturbative EFT corrections to two-derivative gravity theories lead to singularities in the tidal deformations near the horizon in such black holes. The tidal deformations to these metrics have associated scaling exponents as a measure of the strength of the tidal forces. In \cite{Horowitz:2023xyl}, it has been found that for extremal Kerr and Reissner-Nordstr\"om black holes, the EFT corrections to these scaling exponents may come out with a negative sign which results in a significant enhancement of the near-horizon tidal forces. Later, Ref. \cite{Horowitz:2024dch} extended these results for Kerr-Newman black holes and showed that the presence of a $U(1)$ charge on a rotating black hole strengthens the diverging tidal deformations by further reducing the value of the scaling exponents. As discussed in \cite{Horowitz:2023xyl,Horowitz:2024dch} as well as \cite{Chen:2024sgx}, this can be interpreted as a breakdown of the EFT at the horizon of extremal black holes.
 
Building upon the results of \cite{Horowitz:2023xyl,Horowitz:2024dch,Chen:2024sgx}, our work serves as another non-trivial investigation on the EFT corrections to an extremal black hole solution. Here, we calculate the scaling exponents of the near-horizon tidal deformations in an EFT-corrected Majumdar-Papapetrou solution. We consider a system of two black holes, where the effect of one of the black holes on the other can be treated perturbatively. This allows us to model the system as an EFT-corrected extremal Reissner-Nordstr\"om solution, with tidal deformations near the horizon caused by the presence of the second black hole in the environment. The EFT correction terms that we consider are four-derivative in nature, constructed by appropriate contractions of the curvature tensors $(R_{abcd}, R_{ab}, R)$ and the electromagnetic field strength tensor $(F_{ab} = \partial_a A_b - \partial_b A_a)$. We find that the EFT corrections to these scaling exponents vanish for $D=4$ and are negative for $D \geq 5$ \footnote{Previously, \cite{Hadar:2017ven} has also studied the breakdown of EFT on the horizons of extremal black holes in the context of Aretakis instability of the linear perturbation equations. We should emphasize here that the singularities being studied here stem from a completely different source.}.

The expressions of the scaling exponents calculated in our work are of the same form as those of \cite{Horowitz:2023xyl}. Therefore, our results serve as an affirmation that the divergence of tidal forces with EFT corrections is a general phenomenon and can be present in a non-trivial setup like the Majumdar-Papapetrou solution, where near-horizon tidal deformations are already present as a part of the full solution.

The plan of the paper is as follows. In Section \ref{sec:horowitzreview}, we briefly review the work \cite{Horowitz:2023xyl} and its formalism, which lays the foundation for this work. In Section \ref{sec:MPsolreview}, we discuss the Majumdar-Papapetrou solution, its connection to the extremal Reissner-Nordstr\"om black holes, and how we incorporate EFT corrections into its near-horizon sector. Section \ref{sec:sols} discusses the algorithm that we use for solving the equations of motion and proceeds to discuss the solutions in different dimensions and some general formulae for the EFT corrections to near-horizon $AdS_2$ length scale, horizon radius and scaling exponents for $4 \leq D \leq 10$. We conclude with a discussion on the future prospects of this work in \ref{sec:Conclusions}. 

\section{EFT corrections to extremal Reissner-Nordstr\"om (RN) black holes: A Review}\label{sec:horowitzreview}

We start with a review of the results obtained in \cite{Horowitz:2023xyl}, which analyzes the effects of higher-derivative corrections on the near-horizon structure of extremal Reissner-Nordstr\"om (RN) black holes. The EFT corrections to the Einstein-Hilbert action considered there consist of four-derivative terms given by
\begin{equation}
    \begin{split}
        \mathcal{L} = &\left( \frac{1}{2\kappa^2} R - \frac{1}{4} F^{ab}F_{ab} \right) + \mathcal{L}_4 \\
        \mathcal{L}_{4} = & d_{1} R^{2} + d_{2} R_{a b} R^{a b} + d_{3} R_{a b c d} R^{a b c d} \\
        & +\kappa^{2}\left(d_{4} R~ F^{2} + d_5 R_{ab} F^{bc} F_c^{~a} + d_6 R_{abcd} F^{ab} F^{cd}\right) \\
        & +\kappa^{4}\left(d_7 (F_{ab}F^{ab})^2 + d_8 F_{ac} F_{b}^{~c} F^{ad} F_{d}^{~b}\right),    
\end{split}
\end{equation}
$F_{ab}=\partial_a A_b- \partial_b A_a$ encodes the gauge field interactions, and the $d_i$s are dimensionless Wilson coefficients, treated as small parameters. Further calculations and analyses are performed up to first order in $d_i$s. 

To demonstrate the effect of these higher-derivative terms, it is advantageous to decompose the near-horizon extremal RN metric into a product of an $AdS_2$ and an $S^{D-2}$ spacetime as 
\begin{equation}
    \begin{split}
        ds^2_{NH} &= \lads^2 \left( - \rho^2 dt^2 + \frac{d\rho^2}{\rho^2} \right) + r_c^2 ~d\Omega_{D-2}^2~~.
    \end{split}
\end{equation}

For general relativity, the parameters $\lads$ and $r_c$ are related to the horizon radius $r_{H}$ as $\lads = \frac{r_H}{D-3}$ and $r_c = r_H$, respectively. This product form of the near-horizon spacetime is also called the Bertotti-Robinson solution, which in itself is another exact solution to the vacuum Einstein-Maxwell equations \cite{Bertotti:1959pf,Robinson:1959ev}. In the presence of higher-derivative terms, the form of the near-horizon metric and the electromagnetic potential remain unchanged. But, the parameters $\lads$ and $r_c$ get modified by EFT corrections as a function of $d_i$s and $r_H$. 

The stability of this near-horizon geometry is then assessed by introducing linearized static scalar perturbations, which could be sourced by distant matter. Owing to the $AdS_2 \times S^{D-2}$ symmetry of the spacetime, these perturbations are decomposed into $AdS_2$ and $S^{D-2}$ harmonics \cite{Kodama:2003kk, Horowitz:2022mly, Horowitz:2023xyl,Chen:2024sgx}.

\begin{equation}
     \begin{split}
        \delta g_{IJ} &= s\, \mathbb{S}_{j} \rho^{\lambda} g_{IJ}, \\
        \delta g_{AB} &= \rho^{\lambda} \left( h_L \mathbb{S}_{j} g_{AB} + h_T \mathbb{S}_{AB}^{j} \right), \\
        \delta F &= d\left( q\, \mathbb{S}_{j} \rho^{\lambda+1} \right) \wedge dt.
    \end{split} 
\end{equation}

Here, \(\mathbb{S}_j\) represents a spherical harmonic on \( S^{D-2} \), while \(\mathbb{S}^j_{AB}\) denotes the traceless part of its second derivative. The quantities \( s, h_L, h_T, \) and \( q \) are constants, and \( \lambda \) is the scaling exponent that ultimately determines the stability of the perturbations. The indices $(I,~J)$ represent the $AdS_2$ sector, and the indices $(A,~B)$ are along the transverse directions. This decomposition reduces the equations of motion (the Einstein-Maxwell equations + Higher-derivative corrections) to a system of linear equations, which are then solved for the scaling exponent \( \lambda \).

Solving for $\lambda$, Ref. \cite{Horowitz:2023xyl} reports that up to $D=11$, the scaling exponents $\lambda$ as functions of the mode number $j$ can be expressed as
\begin{equation}
    \lambda(j) = -1 + \frac{j}{D-3} - d_0 \frac{8 \kappa^2 (D-4) j (j+D-3)}{(D-2)(D-3)^2 (2j - D + 3) r_H^2}, 
\end{equation}
where the Wilson coefficient combination \( d_0 \) is given by
\begin{equation}\label{horowitzd0}
    \begin{aligned}
        &d_0 =  \frac{(D-3)(D-4)^2}{4} d_1 + \frac{(D-3)(2D^2 - 11D + 16)}{4} d_2 \\
        & + \frac{(2D^3 - 16D^2 + 45D - 44)}{2} d_3 \\
        & + \frac{(D-3)(D-2)(D-4)}{2} d_4  + \frac{(D-2)(D-3)^2}{2} (d_5 + d_6) \\
        &+ \frac{(D-2)^2 (D-3)}{2} (2 d_7 + d_8).
    \end{aligned}  
\end{equation}
Notably, this correction vanishes in \(D=4\), but in higher dimensions, it introduces a shift in the exponent \(\lambda(j)\). The combination of Wilson coefficients $d_0$ is actually a linear combination of the $d_i$s, which keeps any physical observable invariant under field redefinitions \cite{Cheung:2018cwt}. 

Now, there are several interesting features to be observed from this result. Even without the presence of EFT corrections, the horizon loses its smoothness for generic perturbations in $D \geq 5$ due to $\lambda(j)$ becoming fractional numbers for generic values of $j$. In the presence of EFT corrections, when $d_0 >0$, the value of $\lambda(j)$ decreases from its value without the EFT corrections. This indicates a decrease in the stability of the solutions \footnote{According to \cite{Horowitz:2022mly}, perturbations with non-integer scaling exponents $0 < \lambda(j) <2$ lead to the formation of curvature singularities at the horizon.}. For the specific values of $j = (D-3)$, the GR contribution to $\lambda(j)$ vanishes, and only the EFT corrections contribute to the value of the scaling exponent. In such cases, with $d_0 > 0$, we see that the scaling exponent becomes negative, implying that the perturbation diverges at the horizon. This diverging tidal deformation suggests a possible breakdown of EFT at the horizon of the extremal RN black hole. 

However, as discussed in \cite{Chen:2024sgx}, there are two different breakdowns involved in this process: a breakdown of metric perturbation theory, and a breakdown of the EFT approach. It has been argued that metric perturbation theory has a lower cutoff energy than EFT. As a result, while climbing up the energy scale, by the time the EFT breaks down, metric perturbation theory has already become invalid. Hence, following \cite{Chen:2024sgx}, what we observe here is a consequence of the breakdown of metric perturbation theory, and it doesn't signal a breakdown of the EFT per se.

We would like to analyze whether similar results hold true for a two-black hole Majumdar-Papapetrou metric.

\section{Majumdar-Papapetrou Solution and its connection to Extremal Reissner-Nordstr\"om solution}\label{sec:MPsolreview}

In this section, we will discuss in detail the Majumdar-Papapetrou (MP) solution, a configuration of multiple charged black holes that is a solution to the Einstein-Maxwell equations. Following a similar discussion in \cite{Chen:2024sgx}, we will elaborate on how it can be modeled perturbatively as an extremal Reissner-Nordstr\"om black hole whose horizon undergoes tidal deformation due to the presence of other black holes in the background. We will also discuss the coordinate transformations that decompose the Majumdar-Papapetrou metric into a product of a perturbed $AdS_2$ and an $S^{D-2}$ spacetime and end with a discussion on how we cast the form of the metric to include the EFT corrections.

\subsection{MP Solution and Reparametrizations}

The Majumdar-Papapetrou solution is a well-known solution to the Einstein-Maxwell equations where multiple point charges of the same sign (i.e., like charges) can be placed in a static equilibrium such that their gravitational attractions exactly cancel their electrostatic repulsion \cite{Majumdar:1947eu,Papaetrou:1947ib,Albacete:2024qja}. First formulated in $4$ dimensions, this class of solutions was later extended to include black holes in higher dimensions \cite{Hartle:1972ya,Myers:1986rx,Lucietti:2020ryg}. The fact that the gravitational interactions exactly cancel out the electrostatic repulsions tells us that the charges and the gravitational masses of all the black holes should be related only by some common proportionality constant, which is only possible if all the black holes in the theory are extremal.

The Majumdar-Papapetrou metric and electromagnetic potential, in $D$ dimensions, can be written as \cite{Chen:2024sgx,Myers:1986rx}

\begin{equation}
    \begin{split}
        ds^2 &= -H^{-2} d\tau^2 + H^{\frac{2}{D-3}} (d \hat \rho^2 + \hat \rho^2 d \Omega_{D-2}^2) \\
        A_\mu dx^\mu &= \frac{1}{\kappa}\sqrt{\frac{D-2}{D-3}}\frac{1}{H} d\tau
    \end{split}
\end{equation}
where $\kappa^2 = 8\pi G$.\\
The function $H$ is given by 
\begin{equation}
    H(\mathbf{x}) = 1 + \sum_{i=1}^N \frac{M_i}{|\mathbf{x}- \mathbf{x}_i|^{D-3}}
\end{equation}
where $N$ is the total number of extremal black holes in the spacetime, $M_i$ and $\mathbf{x}_i$ are the mass and the position of the horizons of the $i^{th}$ black hole, respectively. In this work, we'll pertain our discussion to the case of only two extremal black holes. Hence, we'll set $N=2$ for the rest of the paper. 

For convenience, we choose the first black hole at $\mathbf{x}_1 = 0 $ as the reference and set $M_1=M$. We denote the position of the second BH as $r_2=\left|\mathbf{x}_2\right|$ and $\hat{\rho}=|\mathbf{x}|$, and defining $\theta$ as the angle that the position vector of the second BH subtends on the radial coordinate as $\mathbf{x} \cdot \mathbf{x}_2=\hat{\rho} ~r_2 \cos \theta$. Therefore, we can write
\begin{equation}
    \begin{split}
        H(\mathbf{x}) &= 1 + \frac{M}{\hat \rho ^{D-3}} + \frac{M_2}{|\hat \rho^2 + r_2^2 + 2 \hat \rho~ r_2 ~\cos\theta|^{D-3}}
    \end{split}
\end{equation}
Near the horizon of the first black hole ($\hat \rho \to 0$), we can expand the denominator of the last term into an infinite series of Gegenbauer polynomials $C_j^\alpha(x)$s. Using this, we can express $H(\mathbf{x})$ as 
\begin{equation}
    \begin{split}
        H(\mathbf{x}) 
        = &\left[ 1 + \frac{M}{\hat \rho^{D-3}} + \frac{M_2}{r_2^{D-3}} C^{\frac{D-3}{2}}_0 (\cos \theta) \right]\\
        &+ \sum_{j=1}^\infty \frac{M_2}{\hat\rho^{D-3}} \left( \frac{\hat \rho}{r_2} \right)^{D-3+j} C^{\frac{D-3}{2}}_j (\cos \theta)\\
     = &\left[ 1 + \frac{1}{\left(\frac{\hat \rho}{r_H}\right)^{D-3}} + \frac{M_2}{r_2^{D-3}} C^{\frac{D-3}{2}}_0 (\cos \theta) \right] \\
     &+ \sum_{j=1}^\infty \frac{M_2}{M} \left(\frac{r_H}{r_2}\right)^{D-3} \left(\frac{\hat \rho}{r_2}\right)^j C^{\frac{D-3}{2}}_j (\cos \theta)
    \end{split}
\end{equation}
where in the second equation, we have used the following relation $M = r_H^{D-3}$ between the mass and the horizon radius of the first black hole.
At this stage, we can define a parameter $\epsilon$ as
\begin{equation}
    \epsilon = \frac{M_2}{M} \left(\frac{r_H}{r_2}\right)^{D-3} = \frac{M_2}{r_2^{D-3}}
\end{equation}
where we see that $\epsilon<<1$ either when the second black hole is much lighter than the first one (i.e., $M_2<<M$) or when it is located at a distance much farther compared to the horizon radius of the first black hole (i.e., $r_2 >> r_H$). At $\hat \rho \to 0$, we can then approximate the behavior of $H(\mathbf{x})$ to be 
\begin{equation}
    H(\mathbf{x}) \approx \frac{1}{\left(\frac{\hat \rho}{r_H}\right)^{D-3}} \left[ 1 + \epsilon \sum_{j=1}^\infty \frac{\hat \rho^{D-3+j}}{r_H^{D-3}~r_2^j} C^{\frac{D-3}{2}}_j (\cos \theta) \right] 
\end{equation}
Further, we reparametrize the $\hat \rho, ~\tau$, and $\theta$ coordinates as follows
\begin{equation}
    \hat \rho^{D-3} \equiv r_H^{D-3} \rho,~~~ (D-3) \tau \equiv t,~~~ \cos \theta \equiv x
\end{equation}
to finally write $H(\mathbf{x})$ and the line element in the Bertotti-Robinson form decoupled as the product of a perturbed $AdS_2$ metric and an $S^{D-2}$ metric as
\begin{equation}\label{Hfunction}
    \begin{split}
        H(\mathbf{x}) &= \frac{1}{\rho} \left[ 1 + \epsilon~ \sum_{j=1}^{\infty} \left( \frac{r_H}{r_2} \right)^j \rho^{1+\frac{j}{D-3}} C^{\frac{D-3}{2}}_j (\cos \theta)  \right] \\
        &\equiv \frac{1}{\rho} \left[ 1 + \epsilon~ \gamma_0(\rho,x) \right].
    \end{split}
\end{equation}
\begin{equation}\label{Dmetric_potential}
    \begin{split}
        ds^2 &= \left(\frac{r_H}{D-3}\right)^2 \bigg[ - \rho^2 \left( 1 - 2\epsilon~\gamma_0(\rho,x) \right) dt^2 \bigg.\\
        &\left. + \frac{d\rho^2}{\rho^2} \left( 1 + \frac{2}{D-3}\epsilon~\gamma_0(\rho,x) \right) \right] \\
        &+ r_H^2 \left[ 1 + \frac{2}{D-3} \epsilon ~\gamma_0(\rho,x) \right] d\Omega_{D-2}^2, \\  
        A_\mu dx^\mu &= \frac{1}{\kappa}\sqrt{\frac{D-2}{D-3}}\left(\frac{r_H}{D-3}\right) ~\rho \left[1 - \epsilon~\gamma_0 (\rho,x) \right] ~ dt.
    \end{split}
\end{equation}
The connection of the Majumdar-Papapetrou solution to the extremal Reissner-Nordstr\"om black hole can be made very clear at this stage by noticing that if we turn off the effect of the second black hole, i.e., $\epsilon = 0$, we recover the familiar form of the metric and the electromagnetic potential as
\begin{equation}
\begin{split}
    ds^2 &= L_{AdS_2}^2 \left(-\rho^2 dt^2 + \frac{d\rho^2}{\rho^2} \right) + r_c^2 d\Omega^2
\end{split}
\end{equation}
\begin{equation}
\begin{split}
    A_\mu dx^\mu &= \frac{1}{\kappa}\sqrt{\frac{D-2}{D-3}}\left(\frac{r_H}{D-3}\right)~\rho~dt
\end{split}
\end{equation}
where we identify $L_{AdS_2} = \frac{r_H}{D-3}$, and $r_c = r_H$.

At this stage, it is already evident from \eqref{Dmetric_potential} that for $D\geq 5$, the tidal deformations are non-smooth \cite{Welch:1995dh,Candlish:2007fh}. Additionally, as mentioned in \cite{Horowitz:2023xyl}, EFT corrections to these tidal deformations further strengthen the divergences. Hence, we will use this setup of the Majumdar-Papapetrou solution for two black holes as an exemplary case where tidal deformations are already present as a part of the full solution and investigate the effect of EFT corrections on the scaling exponents of these tidal deformations.

\subsection{EFT Corrections to MP solution}

We will work with the EFT correction terms as in \cite{Horowitz:2023xyl}, hence the total Lagrangian density would look like
\begin{equation}\label{actionMP}
\begin{split}
    \mathcal{L} = &\frac{1}{2\kappa^2} R  -  \frac{1}{4} F^{ab} F_{ab}  \\
    &+ \bigg[ \frac{1}{\kappa^2} \left( d_1 R^2 + d_2 R_{ab} R^{ab} + d_3 R_{abcd} R^{abcd} \right) \bigg. \\
    &+ \left( d_4 R F^{ab}F_{ab} + d_5 R_{ab} F^{bc} F_c^{~a} + d_6 R_{abcd} F^{ab} F^{cd} \right) \\
    &+ \bigg. \kappa^2 \left( d_7 (F_{ab}F^{ab})^2 + d_8 F_{ac} F_{b}^{~c} F^{ad} F_{d}^{~b} \right) \bigg].
\end{split}
\end{equation}

where we are now using rescaled Wilson coefficients $d_i$s which have a dimension of $[L]^{2}$ as in \cite{Horowitz:2024dch}. 

The equations of motion to be solved are given by appropriate variations of the $\mathcal{L}$ with respect to the metric and the electromagnetic potential, respectively. We can construct an ansatz form for the EFT-corrected Majumdar-Papapetrou solution by introducing appropriate $d_i$ corrections in the length scales and the metric coefficients as

\begin{widetext}
\begin{equation}\label{eftmetansatz}
    \begin{split}
        ds^2 = &\left( \frac{r_H}{D-3}+ d_i ~l_{1i} \right)^2 \left[ - \rho^2 dt^2 \left(1-2\epsilon \gamma^{(t)}_i  \right) + \frac{d\rho^2}{\rho^2} \left( 1 + \frac{2}{D-3} \epsilon \gamma^{(\rho)}_i  \right)\right] \\
        &+ (r_H + d_i~ l_{2i})^2
        \Bigg[
        \frac{dx^2}{1 - x^2} \left( 1 + \frac{2}{D-3} \epsilon \gamma^{(x)}_{i} \right) 
        + (1 - x^2) d\phi_3^2 \left( 1 + \frac{2}{D-3} \epsilon \gamma^{(\phi_3)}_{i} \right) \\
        &\quad + (1 - x^2)\sum_{k=4}^{D} \left( \prod_{l=3}^{k-1} \sin^2\phi_l \right) d\phi_k^2 
        \left( 1 + \frac{2}{D-3} \epsilon \gamma^{(\phi_k)}_{i} \right)
        \Bigg].
    \end{split}
\end{equation}

and the corresponding EFT-corrected electromagnetic potential is given by \footnote{We have seen that for any value of $l_3$, there exists a consistent solution for $l_1$ and $l_2$, and $l_3$ stays free throughout. Hence, we set $l_3$ to $0$ in our further calculations.}
\begin{equation}\label{Amuansatz}
    A_\mu dx^\mu = \frac{1}{\kappa}\sqrt{\frac{D-2}{D-3}}\left(\frac{r_H}{(D-3)} + d_i~l_{3i}\right)  \rho \bigg[1 - \epsilon \gamma_i^{(A)} \bigg]~dt
\end{equation}
\end{widetext}
The corrections in \eqref{eftmetansatz} and \eqref{Amuansatz} have been expressed in short as $d_i ~ l_i = \sum_{i=1}^8 d_i l_i$. The near-horizon symmetry constrains all the correction terms at $\mathcal{O}(\epsilon^1 d_i^1)$ to be of the form
\begin{equation}
\begin{split}
    \gamma^{(a)}(\rho,x) &= \left( \frac{r_H}{r_2} \right)^j \rho^{\left(1+\frac{j}{D-3}\right)+ d_in_i} \left(C^{\frac{D-3}{2}}_{j}(x) + d_i Q^{(a)}_i(x)\right) \\
\end{split}
\end{equation}
where, in $\gamma^{(a)}$s, $(a)$ denotes the component of the metric that gets corrected, and $Q^{(a)}_i(x)$ represent the corrections to the transverse sector, expressed as linear combinations of Gegenbauer polynomials and their derivatives. The sign of $n_i$ indicates how the tidal deformations are affected by the EFT correction terms.

\section{Solutions to EFT Corrections and Tidal Deformations near the horizon}\label{sec:sols}

In this section, we discuss the solutions of the $\gamma^{(a)}(\rho,x)$s and the $n_i$s that solve the Einstein-Maxwell system of equations. 
We start with a discussion of the general algorithm that we have used for solving the $\gamma^{(a)}$s and $n_i$s, which can be extended to all dimensions.
For $4$ dimensions, although some of the $\gamma_a(\rho,x)$ can have non-zero values, it turns out that all the $n_i$s vanish. 
To investigate the behavior of $n_i$s in higher dimensions, we first calculate the same quantities for $ D = 5$ and then extend our calculations up to $ D = 10$. In the end, we discuss some general formulae that fit the results of the EFT corrections to $\lads, r_c$, and $n_i$.

\subsection{General Algorithm}\label{subsec:algorithm}

\noindent Let us now briefly discuss the algorithm that we have adopted to solve the equations of motion at \(\mathcal{O}(\epsilon \, d_i)\).

\noindent Before we start, we would like to point out that while a direct harmonic decomposition of the perturbation as discussed in \cite{Horowitz:2023xyl, Chen:2024sgx} would simplify the analysis due to the high degree of symmetry in the background, we have still chosen to present this algorithmic approach for the following pedagogical reasons. First, it makes the structure of the equations and our notation transparent for later use. Second, since our computations build upon the background metric of \cite{Chen:2024sgx}, this approach proved more natural within our framework.

\noindent We solve for two different sets of equations of motion that follow from the \eqref{actionMP} as: 

\begin{subequations}
  \label{eq:system}
  \begin{align}
    &E^F_b = \frac{\delta \mathcal{L}}{\delta A^b} = 0 \nonumber \\
    \label{eq:EOMA}
    &\Rightarrow  \nabla^a F_{ab} = - \left( d_{1,2,3} \times 0 +~ d_{4,5,6} J^{d_{4,5,6}}_b + \kappa^2 d_{7,8} J^{d_{7,8}}_b \right), \\
    &E^{G}_{ab} = \frac{\delta \mathcal{L}}{\delta g^{ab}} = 0 \nonumber \\
    \label{eq:EOMg}
    &\Rightarrow  G_{ab} - \kappa^2\left(F_a^c F_{bc}-\frac{1}{4}g_{ab}F_{cd}F^{cd}\right) \nonumber \\
    &= -\left( d_{1,2,3} T^{d_{1,2,3}}_{ab} + \kappa^2 d_{4,5,6} T^{d_{4,5,6}}_{ab} + \kappa^4 d_{7,8} T^{d_{7,8}}_{ab} \right).
  \end{align}
\end{subequations}

Here, the current \( J^{d_i}_b \) and the stress-energy tensor \( T^{d_i}_{ab} \) are obtained by varying the EFT correction terms in the action with respect to the gauge field \( A^b \) and the metric \( g^{ab} \), respectively.

Equation \eqref{eq:EOMg} represents a system of coupled differential equations governing the gravitational dynamics. The diagonal components—\((t,t)\), \((\rho,\rho)\), and \((\phi_k,\phi_k)\)—describe the evolution of the system, where the index \( k \) runs from \( 3 \) to \( D \), accounting for the \( D-3 \) angular directions \( \phi_k \). We find that the differential equations corresponding to each \( \phi_k \) are identical.

The off-diagonal components of \eqref{eq:EOMg} and the diagonal component \((x,x)\) impose additional constraints on the system. In particular, the \((\rho,\phi_k)\) and \((x,\phi_k)\) equations enforce the equality of all functions \( Q_i^{(\phi_k)}(x) \), further simplifying the structure of the equations. Taking these constraints into account, the number of independent second-order coupled differential equations reduces to three. Along with this, we find from \eqref{eq:EOMA} that only the timelike component gives us another second-order coupled differential equation of the \(Q^{(a)}(x)\)'s, whereas all the other components are identically zero. 
We thus end up with four independent second-order coupled differential equations and two coupled first-order differential equations. 

To decouple, we choose the following set of equations \(E^F_t = 0\), \(E^G_{tt} + E^G_{\rho\rho}=0\), \(E^G_{\theta\theta} - E^G_{\phi_3\phi_3}=0\), \(E^G_{\rho\rho} - \left(\frac{D-3}{D-4}\right)E^G_{\phi_3\phi_3} = 0\) and \(E^G_{\rho\theta} = 0\), where the first four are second-order ODEs, and the last one acts as constraint equation of the following functions - \(Q^{(t)}_i, Q^{(\rho)}_i, Q^{(x)}_i, Q^{(\phi_3)}_i\, \text{and}\, Q^{(A)}_i\).
We observe at this step that it becomes imperative to decompose these functions in terms of symmetric and anti-symmetric blocks given by : 
\begin{equation}
\begin{split}
     &Q^{(t)}_i=\frac{F^{(2)}_i(x)+F^{(1)}_i(x)}{2(D-3)} \; ; \; Q^{(\rho)}_i=\frac{F^{(2)}_i(x)-F^{(1)}_i(x)}{2} \\
     & Q^{(x)}_i=\frac{F^{(3)}_i(x)+F^{(4)}_i(x)}{2} \; ; \; Q^{(\phi_3)}_i=\frac{F^{(4)}_i(x)-F^{(3)}_i(x)}{2(D-3)}
\end{split}
\end{equation}

On writing our set of differential equations in terms of \(Q^{(A)}_i \) and these new variables \(F^1_i(x), F^2_i(x), F^3_i(x), F^4_i(x)\), \(E^F_t = 0\) reduces to the standard differential equation for $C^{\left(\frac{D-3}{2}\right)}_j(x)$ with some source terms, from which it becomes evident that \(Q^{(A)}_i \) must be of the form $S_A C^{\left(\frac{D-3}{2}\right)}_j(x)$, where $S_A$ is an undetermined constant.
Using this back in the equation \(E^F_t = 0\), \(E^G_{tt} + E^G_{\rho\rho}=0\) and \(E^G_{\rho\theta} = 0\) allows us to determine 
\begin{equation}\label{Ffinalform}
\begin{aligned}
    F^{(1)}_i(x) &= S_1 C^{\left(\frac{D-3}{2}\right)}_j(x), \quad~~~~
    F^{(2)}_i(x) = S_2 C^{\left(\frac{D-3}{2}\right)}_j(x), \\
    F^{(3)}_i(x) &= S_3 \frac{d}{dx} C^{\left(\frac{D-3}{2}\right)}_{j+1}(x), \quad
    F^{(4)}_i(x) = S_4 C^{\left(\frac{D-3}{2}\right)}_j(x)
\end{aligned}
\end{equation}

where all the constants \(S_1,S_2,S_3,S_4\) can be written in terms of $S_A$. 

Substituting \eqref{Ffinalform} in \(E^G_{\rho\rho} - \left(\frac{D-3}{D-4}\right)E^G_{\phi_3\phi_3} = 0\) reduces it into an algebraic equation whence we can solve for $n_i$.

At this point, $S_A$ is still an undetermined constant, and we see that even without fixing its value, we can find the solutions for $n_i$s. Studying the explicit solutions in $D\geq 5$, we see that $Q^{(3)}_i(x)$ and $Q^{(4)}_i(x)$ lead to conical singularities at $x = \pm 1$ due to extra powers of $(1-x^2)$ in the denominator. Hence, to remove these conical singularities, one needs to set such terms to zero. Thus, $Q^{(3)}_i(x)=Q^{(4)}_i(x)=0$ serves as the condition for fixing $S_A$, and we see that both of them lead to the same solution of $S_A$. For $D=4$ dimensions, the extra $(1-x^2)$ factors, as in the higher dimensions, are absent, but to maintain uniformity, we follow the same procedure to fix $S_A$. We emphasize that this does not affect the evaluation of the scaling exponents.

\subsection{Results for $4D$}\label{subsec:4dresults}

Following the algorithm discussed in \ref{subsec:algorithm}, we first calculate the EFT corrections to $\lads$ and $r_c$ as
\begin{equation}
    \begin{split}
        \lads^{(1)} &= \frac{2}{r_H} \left( 2~d_4 + d_5 +2~d_6 +4~d_7 + 2~d_8 \right) \\
        r_c^{(1)} &= \frac{4}{r_H} \left( d_4 - 2~d_7 - d_8 \right) 
    \end{split}
\end{equation}
Next, we proceed to calculate the EFT corrections to the near-horizon tidal deformations as outlined in \ref{subsec:algorithm}. In $D=4$ dimensions, the relevant Gegenbauer polynomials are given by $C^\frac{1}{2}_j(x) = P_j(x)$, which are actually the Legendre polynomials of the first kind. Simultaneously, we will also solve for the EFT corrections to the scaling exponents of the tidal deformations. Solving explicitly for all of these quantities, we find

\begin{table}[h]
    \centering
    \renewcommand{\arraystretch}{1.8} 
    \setlength{\tabcolsep}{12pt} 
    \begin{tabular}{|c|c|c|c|}
        \hline
        $d_i$ & $Q^{(t)}$ & $Q^{(\rho)}$ & $Q^{(A)}$ \\
        \hline
         $d_4$ & $16 (2 + j) $ & $0$ & $8 (2 + j) $ \\ 
         $d_5$ & $4 (1 + j) $ & $-4 $ & $-4 $ \\ 
         $d_6$ & $-8 $ & $-8 $ & $-8 (3 + j) $ \\
         $d_7$ & $-32 $ & $-32 $ & $-16 (4 + j) $ \\ 
         $d_8$ & $-16 $ & $-16 $ & $-8 (4 + j) $ \\ 
        \hline
        \end{tabular}
    \caption{Table of non-zero solutions for $Q^{(t)}$, $Q^{(\rho)}$, and $Q^{(A)}$ in $4D$ for different $d_i$ values in units of $\frac{(r_H/r_2)^j}{r_H^2} C^\frac{1}{2}_j(x)$.}
\end{table}
and, 
\begin{equation}
    n_i = 0 \text{   , for all } i \in [1,8]
\end{equation}
Hence, for $D=4$ dimensions, we find that there are no EFT corrections to the scaling exponents of near-horizon tidal deformations in Majumdar-Papapetrou solutions. This implies that the smoothness or the stability of the horizon is not affected by the EFT corrections. Our results for the $n_i$s are the same as those obtained in \cite{Horowitz:2023xyl} for $D =4$ extremal RN black holes, and this motivates us further to extend this analysis to higher dimensions.

\subsection{Results for $5D$}

Following suit for $D=5$ dimensions, we first solve for the EFT corrections to $\lads$ and $r_c$.

\begin{widetext}
    \begin{equation}
    \begin{split}
        \lads^{(1)} = \frac{1}{r_H} &\left[ \frac{ 11~d_1 + 13~d_2 + 29~d_3 }{6} \right.
        + \bigg. \left( 7~d_4 + 5~d_5 + 10~d_6 + 18~d_7 + 9~d_8 \right) \bigg]  \\
        r_c^{(1)} = \frac{4}{r_H} &\left[ \frac{2~d_1 + 4~d_2 + 8~d_3}{3} +  \left( 2~d_4 + d_5 + 2~d_6 \right)\right] 
    \end{split}
\end{equation}
\end{widetext}

Unlike in $D=4$, here, there are non-zero contributions to $\lads$ and $r_c$ even from purely curvature-squared EFT correction terms. This is evident from the presence of $d_1,~d_2,$ and $d_3$ in the expressions.

The EFT corrections to the near-horizon metric, the electromagnetic potential, and the scaling exponents of tidal deformations can be calculated to be

\begin{widetext}

\begin{table}[h]
    \centering
    \renewcommand{\arraystretch}{1.8} 
    \setlength{\tabcolsep}{12pt} 
    \begin{tabular}{|c|c|c|c|c|}
        \hline
        $d_i$ & $Q^{(t)}$ & $Q^{(\rho)}$ & $Q^{(A)}$ \\
        \hline
        $d_1$ & $(67 + 46 j + 8 j^2 )$ & $-2 (5 + 2 j) $ & $\frac{(108 + 7 j (11 + 2 j))}{3 }$ \\  
        $d_2$ & $(53 + 26 j + 4 j^2) $ & $-2 (19 + 10 j) $ & $\frac{(108 + 5 j (11 + 2 j))}{3 }$ \\  
        $d_3$ & $(109 + 50 j + 8 j^2) $ & $-2 (59 + 30 j) $ & $\frac{(252 + j (131 + 26 j))}{3 }$ \\  
        $d_4$ & $6 (31 + 22 j + 4 j^2) $ & $-12 (5 + 2 j) $ & $2 (36 + j (29 + 6 j)) $ \\  
        $d_5$ & $6 (5 + j (4 + j)) $ & $-12 (7 + 3 j) $ & $-4 (9 + 4 j) $ \\  
        $d_6$ & $-12 (5 + 2 j) $ & $-24 (5 + 2 j) $ & $-4 (54 + j (31 + 4 j)) $ \\  
        $d_7$ & $-36 (5 + 2 j) $ & $-72 (5 + 2 j) $ & $-12 (36 + j (19 + 2 j)) $ \\  
        $d_8$ & $-18 (5 + 2 j) $ & $-36 (5 + 2 j) $ & $-6 (36 + j (19 + 2 j)) $ \\ 
        \hline
        \end{tabular}
    \caption{Table of solutions for $Q^{(t)}$, $Q^{(\rho)}$, and $Q^{(A)}$ in $5D$ for different $d_i$ values in units of $\frac{(r_H/r_2)^j}{r_H^2(3+j)} C^1_j(x)$.}
\end{table}

\begin{equation}
\begin{split}
    \lambda(j) = \left( 1 + \frac{j}{2} \right) -\frac{j (j+2)}{ (j+3) r_H^2} &\bigg[\frac{d_1+11 d_2+31 d_3}{6} + \bigg. d_4+2~ (d_5+ d_6)+ 3 ~(2 d_7+ d_8)\bigg]    
\end{split}
\end{equation}
   
\end{widetext}

It is worth noticing here that in contrast with the case in $D=4$ dimensions, the scaling exponent $\lambda(j)$ receives non-zero EFT corrections with a negative sign (assuming all the $d_i$s to be positive). While this behavior aligns with that found in \cite{Horowitz:2023xyl}, there's another interesting feature in our solutions that differs from \cite{Horowitz:2023xyl}. In our solution, the leading order term of $\lambda(j)$  (i.e., with the higher-derivative terms turned off) is always positive ($\geq 1$) for $j\geq 0$. Hence, although the EFT corrections enhance the tidal deformations by reducing the value of the scaling exponent, they can never completely dominate over the leading order behavior as was seen in \cite{Horowitz:2023xyl}. However, since tidal forces and their scaling exponents are physical observables, the resultant divergences in tidal forces on the horizon due to their non-smoothness can also be interpreted as a version of the breakdown of EFT \cite{Horowitz:2023xyl}.

This difference in behavior between our results and those obtained in \cite{Horowitz:2023xyl} can be attributed to the following fact \footnote{We would like to thank Maciej Kolanowski for pointing this out.}. As derived in \cite{Horowitz:2022mly,Horowitz:2022leb}, one can obtain four possible values for the scaling exponents, of which two are eliminated due to their divergent behaviors at the horizon. Among the other two, \cite{Horowitz:2023xyl} works only with the term containing the leading contribution in the tidal divergences. In the case of MP black holes, this case doesn't arise. Instead, while deriving the two-derivative contribution to the scaling exponents starting from the full solutions, one ends up with only the last surviving solution, which is always a positive non-zero number. Furthermore, the equations for the EFT contributions to the scaling exponents depend on their values for the two-derivative contributions. Hence, the total scaling exponent $\lambda(j)$ doesn't depart from this non-vanishing branch, and the possibly divergent $\lambda(j)$s as in \cite{Horowitz:2023xyl} do not come into play at all. This is one of the main results of our paper, that even if one takes into account the EFT corrections to the near-horizon tidal deformations in MP black holes, the scaling exponents still stay well-behaved, and no such case arises where the two-derivative contributions vanish, and hence, the EFT contributions dominate over them. We discuss more about this issue in \ref{sec:Conclusions}

Next, we repeat this algorithm in higher dimensions $(D \leq 10)$ \footnote{We have performed these calculations using the xAct suite of packages \cite{Martin-Garcia:xAct, Brizuela:2008ra,Nutma:2013zea}} to investigate whether we can find a general expression for the EFT corrections to the scaling exponent.

\subsection{Formula for EFT Corrections to $\lads, r_c$ and $n_i$s for up to 10 dimensions}\label{subsec:genscale}

Proceeding with the similar calculations as in the previous subsection, the EFT corrections to $\lads$ and $r_c$ can be expressed for $4 \leq D \leq 10$ as:

\begin{widetext}

\begin{align}
r_H ~\lads^{(1)} = 
&  \frac{(D-4)}{(D-3)(D-2)} \left[ \left(2 D^2-11 D+16\right) ~d_1 + \left(2 D^2-10 D+13\right) ~d_{2} + \frac{2 \left(2 D^3-16 D^2+44 D-41\right)  }{(D-3)}~d_{3} \right] \notag\\
&+ \frac{2 \left(2 D^2-13 D+22\right) }{D-3}~d_{4}  + (3 D-10) (d_{5}+2  ~d_{6})+\frac{2 (2 D-7) (D-2) }{D-3}(2~d_{7}+~d_{8})
\end{align}

\begin{align}
r_H~r_c^{(1)} = 
&  \frac{(D-4)}{(D-2)} \left[ \left(D^2-5 D+8\right) ~d_1 + \left(2 D^2-9 D+11\right) ~d_{2} + \frac{2 \left(2 D^3-16 D^2+45 D-43\right) }{(D-3)}~d_{3} \right]\notag \\
&+ 2 \left(D^2-7 D+14\right) ~d_{4} + 2 (D-4) (D-3) (d_{5}+2~d_{6} ) + 2 (D-5) (D-2) (2~d_{7} + ~d_{8})
\end{align}
\end{widetext}

Furthermore, we find that in all these higher-dimensional cases, the conical singularity in the metric corrections can only be avoided by setting all the angle-dependent transverse metric corrections to zero. Imposing this condition and solving for the equations of motion, $\lambda(j)$ can be expressed for $4 \leq D \leq 10$ by a general formula

\begin{widetext}

\begin{equation}
\begin{split}
    \lambda(j) = 1+\frac{j}{D-3} - \frac{d_0}{r_H^2} \left(\frac{8 (D-4) }{(D-3)^2 (D-2) } \frac{~j~ (j+D-3)}{(2j+3(D-3))} \right) ,
\end{split}
\end{equation}

\noindent where

\begin{equation}
\begin{split}
    d_0 =~ & \frac{(D-3) (D-4)^2}{4} ~ d_1+ \frac{(D-3)\left(2 D^2-11 D+16\right)}{4}   ~d_2+\frac{\left(2 D^3-16 D^2+45 D-44\right)}{2} ~ d_3\\
    &+\frac{(D-3) (D-2) (D-4)}{2}  ~d_4+\frac{(D-3)^2 (D-2)}{2}  ~(d_5+ d_6) \\
    &+ \frac{(D-3) (D-2)^2 }{2}~ (2 d_7+ d_8).
\end{split}
\end{equation}
is the same field-redefinition invariant combination of Wilson coefficients as in \eqref{horowitzd0}.
   
\end{widetext}

Regarding the behavior of the scaling exponent, we see that similar to the case in $D=5$ dimensions, the EFT corrections strengthen the tidal deformations near the horizon, but unlike the case for the extremal RN solution, cannot dominate over the leading order terms. This indicates that even in higher dimensions, the breakdown of EFT, in the sense of $\lambda < 0$, is not observed in the MP solution.

\section{Conclusions}\label{sec:Conclusions}

Our study builds upon the work of \cite{Horowitz:2023xyl}, extending the analysis to examine how EFT corrections modify the two black hole Majumdar-Papapetrou solution of general relativity. This serves as an illustrative case of a system where the horizon is already tidally deformed due to a well-defined source, i.e., the second black hole. In contrast with previous works like \cite{Horowitz:2023xyl} where the source of the tidal deformation was arbitrary, in our case, it is fixed to be of a particular form due to the full solution, i.e., the MP metric.

To begin, we note that the scaling exponent for a Majumdar-Papapetrou black hole is given by $\lambda(j)=1+\frac{j}{D-3}$ \cite{Chen:2024sgx}. As a result, while the horizon remains smooth in $D = 4$, it generally lacks smoothness in higher dimensions \cite{Welch:1995dh,Candlish:2007fh}. After incorporating EFT corrections, we find that for cases when $j < (D-3)$, the scaling exponent satisfies $1 <\lambda(j) < 2$, for all $D>4$, indicating possible divergences in the tidal forces at the horizon, despite the scalar curvatures remaining finite. Furthermore, we observe that the EFT-induced corrections to the scaling exponents, $n_i$s, are zero for $D = 4$ and assume negative values for $5 \leq D \leq 10$. This suggests that although the Majumdar-Papapetrou black hole maintains a smooth horizon at $D = 4$, its smoothness diminishes progressively for $D \geq 5$, consistent with the conclusions of \cite{Horowitz:2023xyl}. 

However, our result differs from that of \cite{Candlish:2007fh} on the differentiability of the horizon for the $j = 1$ mode in $D = 5$. While they report the horizon to be $C^2$ continuous for this mode, it seems to be only $C^1$ in our case. Our result agrees with \cite{Kodama:2003kk,Chen:2024sgx} regarding this particular mode not being absorbable into the metric as a gauge, and hence, being a gauge-invariant physical mode. Nevertheless, note that the $j = 1$ mode does not allow for the construction of symmetric, traceless rank-2 tensor harmonics on $S^3$. Hence, although it remains a gauge-invariant physical mode, it cannot source large corrections to the curvature near the horizon.

Another important point to ponder is the possibility of the excitation of the possibly problematic modes when we incorporate the EFT corrections into the near-horizon MP metric. In our work, we have made assumptions regarding the smallness of the tidal deformations and the EFT corrections that allow us to bypass this issue. We have assumed the EFT corrections to be sufficiently small so that the boundary conditions near the horizon remain unchanged. Furthermore, we have assumed the tidal deformations due to the second black hole to be small enough to consider the background solution in GR as an extremal Reissner-Nordstr\"om solution in the Bertotti-Robinson form, which allows a decomposition into the $AdS_2 \times S^{D-2}$ harmonics. As a consequence of the first assumption, although we analyze the near-horizon structure of the Majumdar-Papaetrou solution (which is known to be valid over the full spacetime in GR) and then consider EFT corrections to it, no new modes are excited due to the boundary conditions remaining unmodified. Due to this assumption of a perturbative setup, upon adding EFT corrections, the equations of motion are linearized in the EFT coupling parameters. This leads to the EFT corrections of the scaling exponents being uniquely determined in terms of their leading-order solutions. Hence, even on taking EFT corrections into account, the other modes are not excited here. 
That said, it is important to point out that our assumption highlights a potential limitation of the present analysis. A definitive exclusion of the problematic mode would require studying the full solution with EFT corrections, rather than restricting attention to the near-horizon region. Moreover, our assumption does not preclude the appearance of such modes in more intricate scenarios. For instance, one could examine whether they arise in the near-horizon analysis of a fully EFT-corrected solution involving multiple black holes. Exploring the emergence of these problematic modes in such systems would certainly be worthwhile.
\footnote{We thank the referee for highlighting these important points.}


Regarding the possibility of the breakdown of EFT, one could adopt two different points of view. Even when $1 <\lambda(j) < 2$, the tidal forces would diverge on the extremal horizon, which can be measured by a physical observer. This should indicate at least a pathology of the EFT, if not a full breakdown. On the other hand, if $\lambda < 0$ (owing to the condition $j=(D-3)$, and $d_0 > 0$), the tidal deformations to the metric also diverge, leading to an amplification of the effects of the EFT terms. Following \cite{Chen:2024sgx}, the latter one is a consequence of the breakdown of metric perturbation theory, which occurs at an even lower energy scale than the EFT cutoff.

It would be an interesting direction of future research to explore how the results discussed in this paper fare for a case of $N$-black hole Majumdar-Papapetrou solution, as studied in \cite{Chen:2024sgx}. 
Another important avenue worth pursuing would be to analyze the effects of the EFT corrections on the tidal deformations in the nonlinear regime. 

\section*{Acknowledgements}

We thank Amitabh Virmani, Maciej Kolanowski, Calvin Y. Chen, Andrew J. Tolley, and Claudia de Rham for their helpful comments on the draft. We would also like to thank Yogesh K. Srivastava for valuable discussions. We would also like to thank the anonymous referee for their insightful comments, which helped us enrich the quality of this manuscript. SS's research is supported by the Department of Science and Technology, Government of India, under the ANRF CRG Grant (No. CRG/2023/000934).

\end{document}